\documentclass[11pt]{article}

\usepackage{scicite}
\usepackage{epsfig}
\usepackage{graphicx}
\usepackage{times}
\usepackage{ccaption}
\usepackage{upgreek}
\usepackage{amsmath}
\usepackage{amssymb}
\usepackage{hyperref}
\usepackage{color}
\usepackage{url}

\newcommand{\apjl}{Astrophys. J. Lett.}
\newcommand{\physrep}{Phys. Rep.}

\newcommand{\araa}{Annu. Rev. Astron. Astrophys.}
\newcommand{\aap}{Astron. Astrophys}

\newcommand{\prd}{Phys. Rev. D}
\newcommand{\prl}{Phys. Rev. Lett.}
\newcommand{\apj}{Astrophys. J.}
\newcommand{\nat}{Nature}

\newcommand{\mdash}{---}

\newenvironment{sciabstract}{%
\begin{quote} \bf}
{\end{quote}}

\title{Measurement of the cosmic-ray proton spectrum from 40 GeV to 100 TeV
with the DAMPE satellite}

\author
{DAMPE Collaboration\footnote{The correspondence should be addressed to: dampe@pmo.ac.cn} :
Q.~An$^{1,2}$, R.~Asfandiyarov$^{3}$, P.~Azzarello$^{3}$,\\
P.~Bernardini$^{4,5}$, X.~J.~Bi$^{6,7}$,
M.~S.~Cai$^{8,9}$, J.~Chang$^{8,9}$, D.~Y.~Chen$^{7,8}$,\\
H.~F.~Chen$^{1,2}$\footnote{Deceased.}, J.~L.~Chen$^{10}$
% ...
}

\date{}

\begin{document}
\baselineskip24pt

\maketitle

\noindent
\normalsize{$^{1}$State Key Laboratory of Particle Detection and Electronics, University of Science and Technology of China, Hefei 230026, China}\\
\normalsize{$^{2}$Department of Modern Physics, University of Science and Technology of China, Hefei 230026, China}\\
\normalsize{$^{3}$Department of Nuclear and Particle Physics, University of Geneva, CH-1211, Switzerland}\\
\normalsize{$^{4}$Dipartimento di Matematica e Fisica E. De Giorgi, Universit\`a del Salento, I-73100, Lecce, Italy}\\
\normalsize{$^{5}$Istituto Nazionale di Fisica Nucleare (INFN) - Sezione di Lecce, I-73100, Lecce, Italy}\\
\normalsize{$^{6}$Institute of High Energy Physics, Chinese Academy of Sciences, Yuquan Road 19B, Beijing 100049, China}\\
\normalsize{$^{7}$University of Chinese Academy of Sciences, Yuquan Road 19A, Beijing 100049, China}\\
\normalsize{$^{8}$Key Laboratory of Dark Matter and Space Astronomy, Purple Mountain Observatory, Chinese Academy of Sciences, Nanjing 210034, China}\\
\normalsize{$^{9}$School of Astronomy and Space Science, University of Science and Technology of China, Hefei 230026, China}\\
\normalsize{$^{10}$Institute of Modern Physics, Chinese Academy of Sciences, Nanchang Road 509, Lanzhou 730000, China}\\
\normalsize{$^{11}$National Space Science Center, Chinese Academy of Sciences, Nanertiao 1, Zhongguancun, Haidian district, Beijing 100190, China}\\
\normalsize{$^{12}$Gran Sasso Science Institute (GSSI), Via Iacobucci 2, I-67100 L'Aquila, Italy}\\
\normalsize{$^{13}$Istituto Nazionale di Fisica Nucleare (INFN) -Laboratori Nazionali del Gran Sasso, I-67100 Assergi, L'Aquila, Italy}\\
\normalsize{$^{14}$Istituto Nazionale di Fisica Nucleare (INFN) - Sezione di Bari, I-70125, Bari, Italy}\\
\normalsize{$^{15}$Istituto Nazionale di Fisica Nucleare (INFN) - Sezione di Perugia, I-06123 Perugia, Italy}\\
\normalsize{$^{16}$Dipartimento di Fisica ``M.~Merlin'' dell'Universit\`a e del Politecnico di Bari, I-70126, Bari, Italy}\\
\normalsize{$^{17}$Department of Physics and Laboratory for Space Research, the University of Hong Kong, Pok Fu Lam, Hong Kong, China}\\

\pagebreak

\baselineskip24pt

\begin{sciabstract}
The precise measurement of the spectrum of protons, the most abundant
component of the cosmic radiation, is necessary to understand the source
and acceleration of cosmic rays in the Milky Way. This work reports the
measurement of the cosmic ray proton fluxes with kinetic energies from
40 GeV to 100 TeV, with two and a half years of data recorded by the
DArk Matter Particle Explorer (DAMPE). This is the first time an experiment directly 
measures the cosmic ray protons up to $\sim 100$ TeV with a high statistics.
The measured spectrum confirms the spectral hardening found by previous 
experiments and reveals a softening at $\sim13.6$ TeV,
with the spectral index changing from $\sim2.60$ to $\sim2.85$.
Our result suggests the existence of a new spectral feature of cosmic
rays at energies lower than the so-called knee, and sheds new light on
the origin of Galactic cosmic rays.
\end{sciabstract}

\section*{Introduction}

It is widely believed that the remnants of explosive stars in the Milky
Way may play a substantial role in producing energetic cosmic ray (CR)
particles\cite{2015ARA&A..53..199G,1934PhRv...46...76B}. The energy spectra
of CRs are expected to be single power-laws (PLs) until the energies exceed the
maximum acceleration limits of the sources, based on the conventional
Fermi acceleration models\cite{1949PhRv...75.1169F}. The diffusive
transportation of CRs in the interstellar turbulent magnetic field
results in a softening of the accelerated spectrum, by again a power-law
form (for rigidities above a few tens of GV) according to the Boron-to-Carbon
ratio\cite{2016PhRvL.117w1102A}. This general picture of CR production
and propagation has been supported by measurements
of CR energy spectra and composition ratios, as well as diffuse
$\gamma$-rays\cite{2007ARNPS..57..285S}.

However, such a simple picture has been challenged by some recent
high-precision measurements. Remarkable spectral hardenings of the
energy spectra of CRs were revealed in the
ATIC\cite{2005ICRC....3..105W,2009BRASP..73..564P},
CREAM\cite{2010ApJ...714L..89A,2009ApJ...707..593A},
PAMELA\cite{2011Sci...332...69A}, and AMS-02\cite{2015PhRvL.114q1103A}
observations. The spectral hardenings suggest extensions of the
traditional CR source injection, acceleration, and/or propagation
processes, e.g., refs.\cite{2012PhRvL.109f1101B,2012ApJ...752...68V,
2013ApJ...763...47P}.
The spectral behaviors of CRs at higher energies ($>$TeV) are
essential to understand the nature of the spectral hardenings, as well
as the origin and propagation of CRs. Moreover, the extension of the
spectra to PeV energies according to the power-law indices measured at
sub-TeV energies seems to be in conflict with the all-particle spectrum
of CRs\cite{2018ChPhC..42g5103G}. Such a puzzle may be solved if a
significant spectral softening presents well below the so-called knee
at several PeV. The precise measurements of the energy spectra of CRs
above TeV are thus motivated by the test of potential new spectral
features. Interestingly, the recent CREAM and NUCLEON data show hints
that the energy spectra of CR nuclei may become softer above rigidities
of $10\sim20$ TV\cite{2017ApJ...839....5Y,2018arXiv180507119A}. However,
the result of the proton plus helium spectrum from the air shower experiment
ARGO-YBJ shows a single power-law form for energies between 3 and 300
TeV\cite{2015PhRvD..91k2017B}. The CREAM result is mainly limited by
its low statistics. The NUCLEON data, with again a relatively low
statistics at energies above tens TeV, also suffer from sizeable
systematic uncertainties that need to be properly included.
The indirect measurements, on the other hand, suffer from poor
composition resolution. Although the magnetic spectrometers can measure
CRs very accurately, they are unable to reach energies well beyond TeV
in the foreseeable future. Therefore, the calorimeter-based direct
measurement experiments, with high statistics up to $\sim 100$ TeV
and well-controlled systematic uncertainties, are most suitable to
solve the above problems.

\section*{Results}

In this work we present the measurement of the proton spectrum with the
Dark Matter Particle Explorer (DAMPE; also known as ``Wukong'' in China).
DAMPE is a calorimetric-type, satellite-borne detector for observations
of high energy electrons, $\gamma$-rays, and CRs\cite{2017APh....95....6C,
ChangJin2014}. From top to bottom, the instrument consists of a
Plastic Scintillator strip Detector (PSD\cite{2017APh....94....1Y}),
a Silicon-Tungsten tracKer-converter (STK\cite{2016NIMPA.831..378A}),
a bismuth germanate BGO imaging calorimeter\cite{2015NIMPA.780...21Z}, and a NeUtron
Detector (NUD\cite{2016AcASn..57....1H}). The PSD measures the charge
of incident particles, and serves as an anti-coincidence detector for
$\gamma$-rays. The STK reconstructs the trajectory and also measures
the charge of the particles. The BGO calorimeter measures the energy
and trajectory of incident particles, and provides effective
electron/hadron discrimination based on the shower images.
The NUD provides additional electron/hadron discrimination.
These four sub-detectors enable good measurements of the charge
($|Z|$) with a resolution (Gaussian standard deviation) of about $0.06e$
and $0.04e$ for the PSD\cite{DAMPE-Charge} and the STK respectively,
the arrival direction with an angular resolution of better than $0.5^{\circ}$
above 5 GeV, the energy with a resolution of higher than $1.5\%$ for $>10$
GeV electrons/photons\cite{2017Natur.552...63D} and about $25\%\sim35\%$
for protons up to 10 TeV\cite{2017APh....95....6C}, and the identification of
incoming particles with a proton rejection capability of about
$3\times10^4$ when keeping $90\%$ of electrons\cite{2017Natur.552...63D}.
The DAMPE detector was launched into a 500-km Sun-synchronous orbit
on December 17, 2015. The on-orbit calibration results demonstrate
that DAMPE operates stably in space\cite{DAMPE-Calibration}.

\begin{figure}[!ht]
\includegraphics[width=0.33\textwidth]{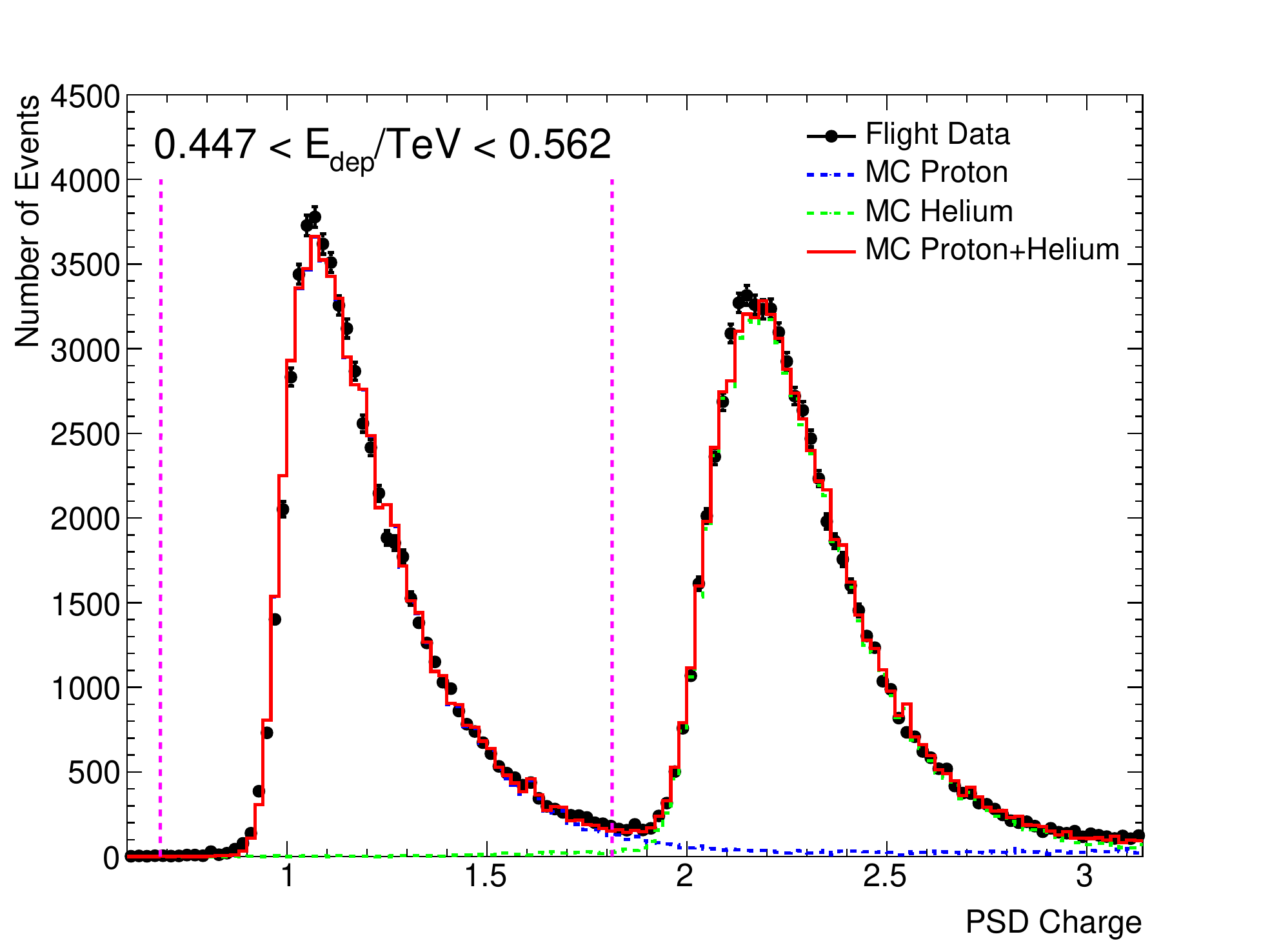}
\includegraphics[width=0.33\textwidth]{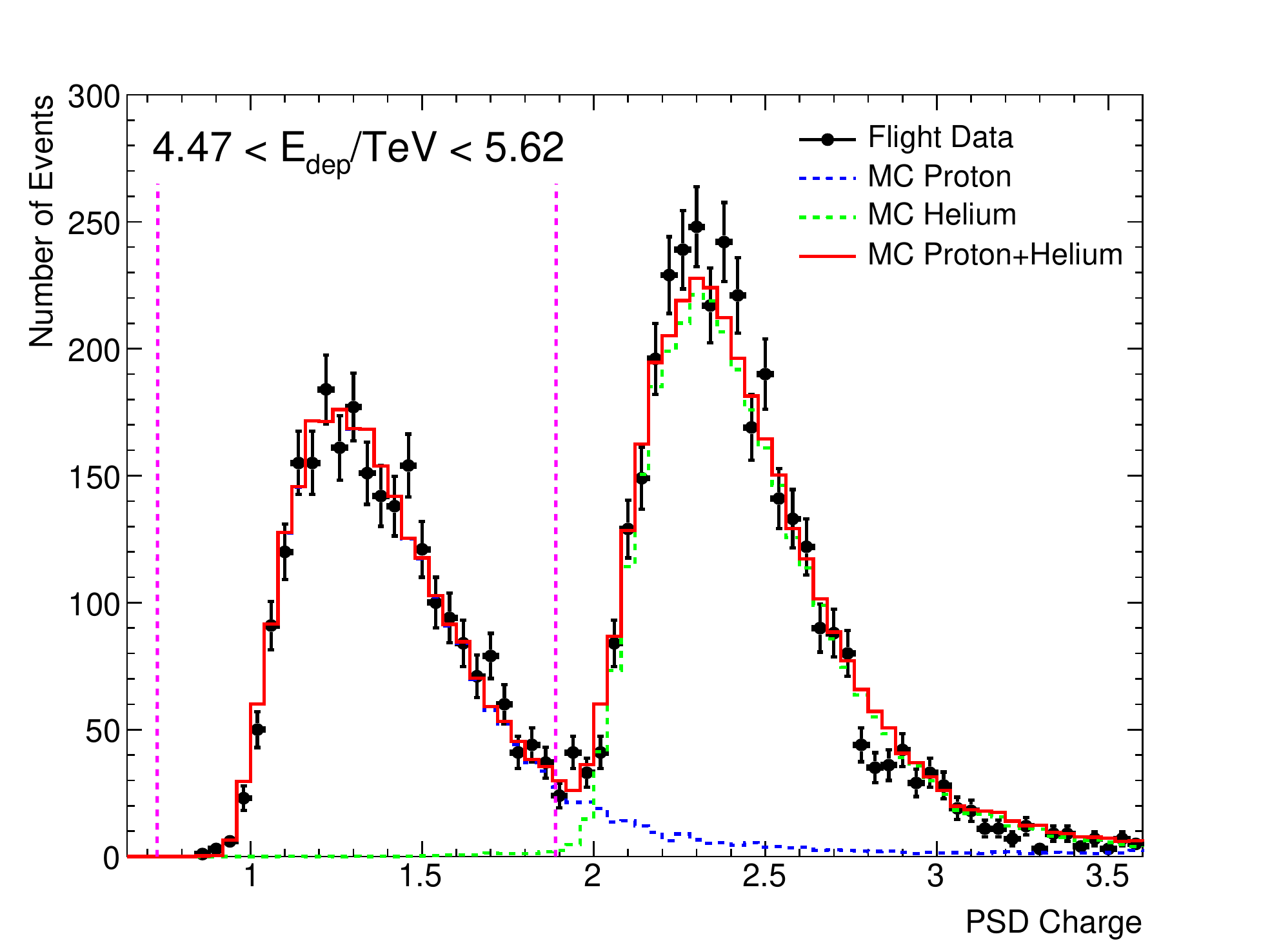}
\includegraphics[width=0.33\textwidth]{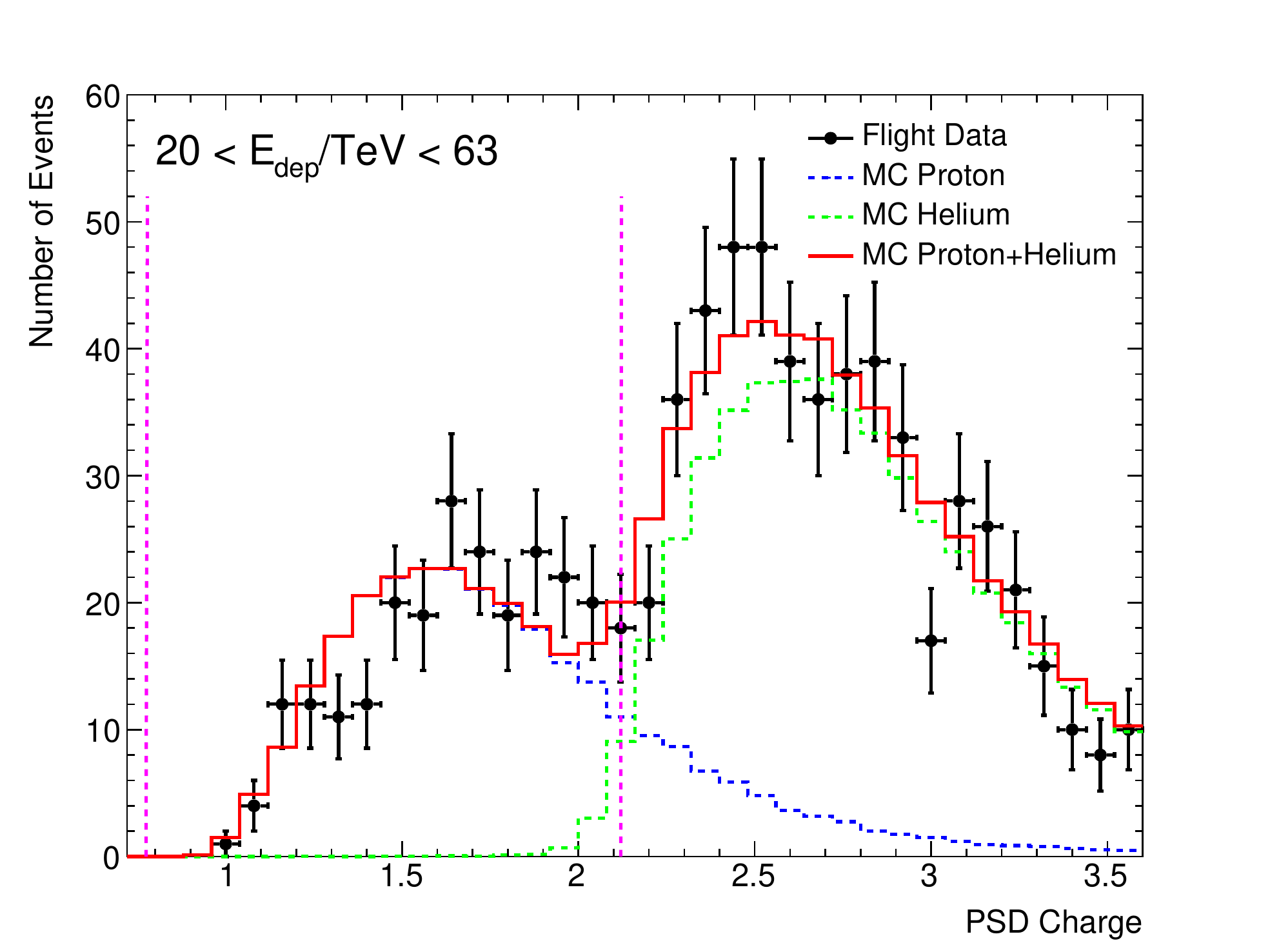}
\caption{{\bf The combined signal spectra of PSD for protons and helium
nuclei.} The left panel is for BGO deposited energies between 447 GeV and 562 GeV,
the middle panel is for BGO deposited energies of $4.47-5.62$ TeV, and
the right panel is for BGO deposited energies between 20 TeV and 63 TeV. The on-orbit
data (black) are shown, together with the best-fit templates of simulations
of protons (blue), helium nuclei (green), and their sum (red). The vertical dashed
lines show the cuts to select proton candidates in this deposited energy range.
}
\label{fig:Figure1}
\end{figure}

The data used in this work cover the first 30 months of operation of
DAMPE, from January 1, 2016 to June 30, 2018. The fraction of live
time is about $75.73\%$ after excluding the time when the satellite passes
the South Atlantic Anomaly region, the instrument dead time, the time for
on-orbit calibration, and the period between September 9, 2017 and September
13, 2017 during which a big solar flare occurred and may have affected the
baseline of the detector. Protons are selected using the charge measured
by the PSD (see the Materials and Methods for details about the event
selection). Figure~\ref{fig:Figure1} illustrates the reconstructed PSD
charge spectra for low-$Z$ nuclei for deposited energies of $447-562$
GeV (left panel), $4.47-5.62$ TeV (middle panel), and $20-63$ TeV
(right panel), together with the Monte Carlo (MC) simulations of protons
and helium nuclei with GEANT v4.10.03\cite{2003NIMPA.506..250A}.
Note that small corrections from the reconstructed charge to the true
particle charge\cite{DAMPE-Charge} have not been applied in this work.
The proton and helium peaks are clearly separated in this plot.
The contamination of the proton sample due to helium nuclei is found to
be less than $1\%$ for deposited energies below 10 TeV and about $5\%$
around 50 TeV, as an effect of the energy-dependent charge selection
(see the Materials and Methods). Given the excellent electron-proton
discrimination capability of DAMPE\cite{2017Natur.552...63D}, the
contamination due to residual electrons is estimated to be about $0.05\%$
in the whole energy range analyzed in this work.

\begin{figure}[!ht]
\includegraphics[width=1.0\textwidth]{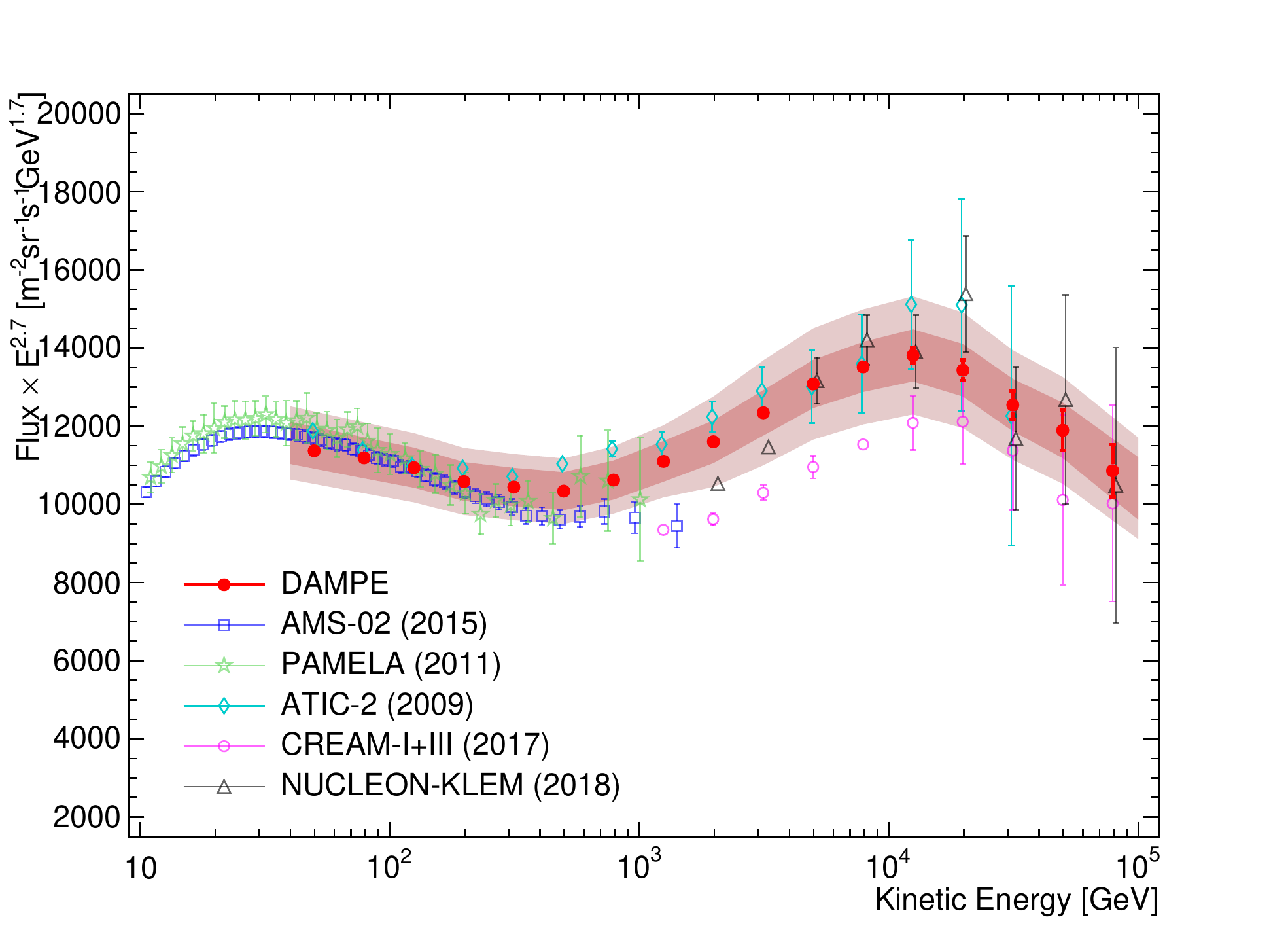}
\caption{{\bf Proton spectrum from 40 GeV to 100 TeV measured with DAMPE
(red filled dots).} The red error bars show the statistical uncertainties,
the inner shaded band shows the estimated systematic uncertainties due
to the analysis procedure, and the outer band shows the total systematic
uncertainties including also those from the hadronic models.
The other direct measurements by PAMELA\cite{2011Sci...332...69A}
(green stars), AMS-02\cite{2015PhRvL.114q1103A} (blue squares),
ATIC-2\cite{2009BRASP..73..564P} (cyan diamonds), CREAM
I+III\cite{2017ApJ...839....5Y} (magenta circles), and NUCLEON
KLEM\cite{2018arXiv180507119A} are shown for comparison. For the
PAMELA data, a $-3.2\%$ correction of the absolute fluxes has been
included\cite{2014PhR...544..323A,PAMELA-10yrs}.
The error bars of PAMELA and AMS-02 data include both statistical and
systematic uncertainties added in quadrature. For ATIC, CREAM, and
NUCLEON data only statistical uncertainties are shown.
}
\label{fig:Figure2}
\end{figure}

The proton spectrum in the energy range from $40$ GeV to $100$ TeV is shown
in Figure~\ref{fig:Figure2} and tabulated in Table 1. Error bars represent
the $1\sigma$ statistical uncertainties of the DAMPE measurements, and the
shaded bands show the systematic uncertainties associated with the analysis
procedure (inner band) and the total systematic uncertainties including those
from the hadronic models (outer band). Previous measurements by space detectors
PAMELA\cite{2011Sci...332...69A} and AMS-02\cite{2015PhRvL.114q1103A},
and balloon-borne detectors ATIC-2\cite{2009BRASP..73..564P},
CREAM\cite{2017ApJ...839....5Y}, and NUCLEON\cite{2018arXiv180507119A}
are overlaid for comparison. The DAMPE spectrum is consistent with those
of PAMELA and AMS-02. At higher energies, our results are also consistent
with that of CREAM, ATIC-2, and NUCLEON when the systematic uncertainties
are taken into account.

\section*{Discussion}

The features of the proton spectrum measured by DAMPE in the energy
range from 40 GeV to 100 TeV give fundamental information about the origin
and propagation of Galactic CRs. A spectral hardening at a few hundred GeV
energies is shown in our data, in agreement with that of
PAMELA\cite{2011Sci...332...69A} and AMS-02\cite{2015PhRvL.114q1103A}.
As discussed in several papers (Ref.\cite{2018mmea.book....1A} and
references therein) the hardening can be due to either details of the
acceleration mechanism, effects in the propagation in the Milky Way,
or the contribution of a new population of CRs (e.g. a nearby source).
Furthermore, the DAMPE measurement gives, for the first time, a strong
evidence of a softening at about 10 TeV.  It is worth reminding that a 
maximum of the large-scale anisotropy has been observed just at that 
energy (see e.g., Ref.\cite{2018mmea.book....1A, 2015ApJ...809...90B}).
We fit the spectrum with energies between 1 TeV and 100 TeV with
a single power-law model and a smoothly broken power-law model respectively,
and find that the smoothly broken power-law model is favored at the $4.7\sigma$
confidence level compared with the single power-law one (see the Materials and Methods
for details of the fit). For the smoothly broken power-law model fit,
the break energy is $13.6^{+4.1}_{-4.8}$ TeV, the spectral index below
the break energy is $2.60\pm0.01$, and the change of the spectral index
above the break energy is $-0.25\pm0.07$.
Results recently published by CREAM\cite{2017ApJ...839....5Y} and
NUCLEON\cite{2018arXiv180507119A} experiments also indicate a spectral
softening at rigidities of $\sim10$ TV. However, these results are
limited by low statistics or the lack of careful studies of
systematic uncertainties.

The spectral hardening and softening are not compatible with the paradigm
of a unique power-law spectrum up to the all-particle knee at PeV energies,
thus implying a deep revision of CR modeling in the Galaxy. For instance
the 10 TeV softening might be due to the exhaustion of the contribution
of a given CR population. Either a local source on top of a power-law
background\cite{2017PhRvD..96b3006L}, or various types of
sources\cite{2006A&A...458....1Z} can be compatible with this scenario.
It should be noted that, the spectral softening should not correspond
to the knee of protons; otherwise the expected all-particle spectrum would
under-shoot the observational data, either for mass-dependent or
charge-dependent knees of various species\cite{2018ChPhC..42g5103G}.
Therefore, the current DAMPE measurement of the proton spectrum,
together with other measurements from the space and groundbased
experiments, puts a severe constraint on the models of Galactic CRs.

\noindent
\section*{Materials and Methods}

\subsection*{MC simulations}

Extensive MC simulations are carried out to estimate the selection
efficiencies, background contaminations, as well as the energy
response matrix of hadronic cascades in the detector (in particular,
the BGO calorimeter). The GEANT v4.10.03\cite{2003NIMPA.506..250A}
is adopted for these simulations. There are two typical hadronic
interaction models in the GEANT simulation tool, the QGSP\_FTFP\_BERT
and FTFP\_BERT models. Comparisons of the shower development
(longitude and transverse distributions) between simulations and the
beam-test and on-orbit data show that the FTFP\_BERT model matches
better with the data. Therefore we adopt the FTFP\_BERT model as the
benchmark one of the MC simulations for incident energies less than
100 TeV. For higher energies we employ the FLUKA tool which links
the DPMJET model for the simulation\cite{2014NDS...120..211B}.
The FLUKA based results are also used as a cross check and an estimate
of the systematic uncertainties of the hadronic interaction models
through comparing with the GEANT results.

An isotropic flux with $E^{-1.0}$ spectrum is generated for the
detector simulation. We simulated protons, helium nuclei, and electrons
in this analysis. After the charge selection, heavier nuclei are negligible
for the proton analysis. In the analysis, the spectra are re-weighted to
$E^{-2.7}$ for protons and helium nuclei, and to $E^{-3.15}$ for electrons.
The final proton spectrum we measured is not exactly the same as $E^{-2.7}$.
However, changing the re-weight spectrum with indices from 2.5 to 3.1 has
little influence on the results.

\subsection*{Proton event selection}

The events with energy deposit in the BGO calorimeter larger than 20
GeV are selected in this analysis in order to suppress the effect of the
geomagnetic rigidity cutoff. The detailed event selection method is
described as follows.

\begin{figure}[!ht]
\includegraphics[width=0.5\textwidth]{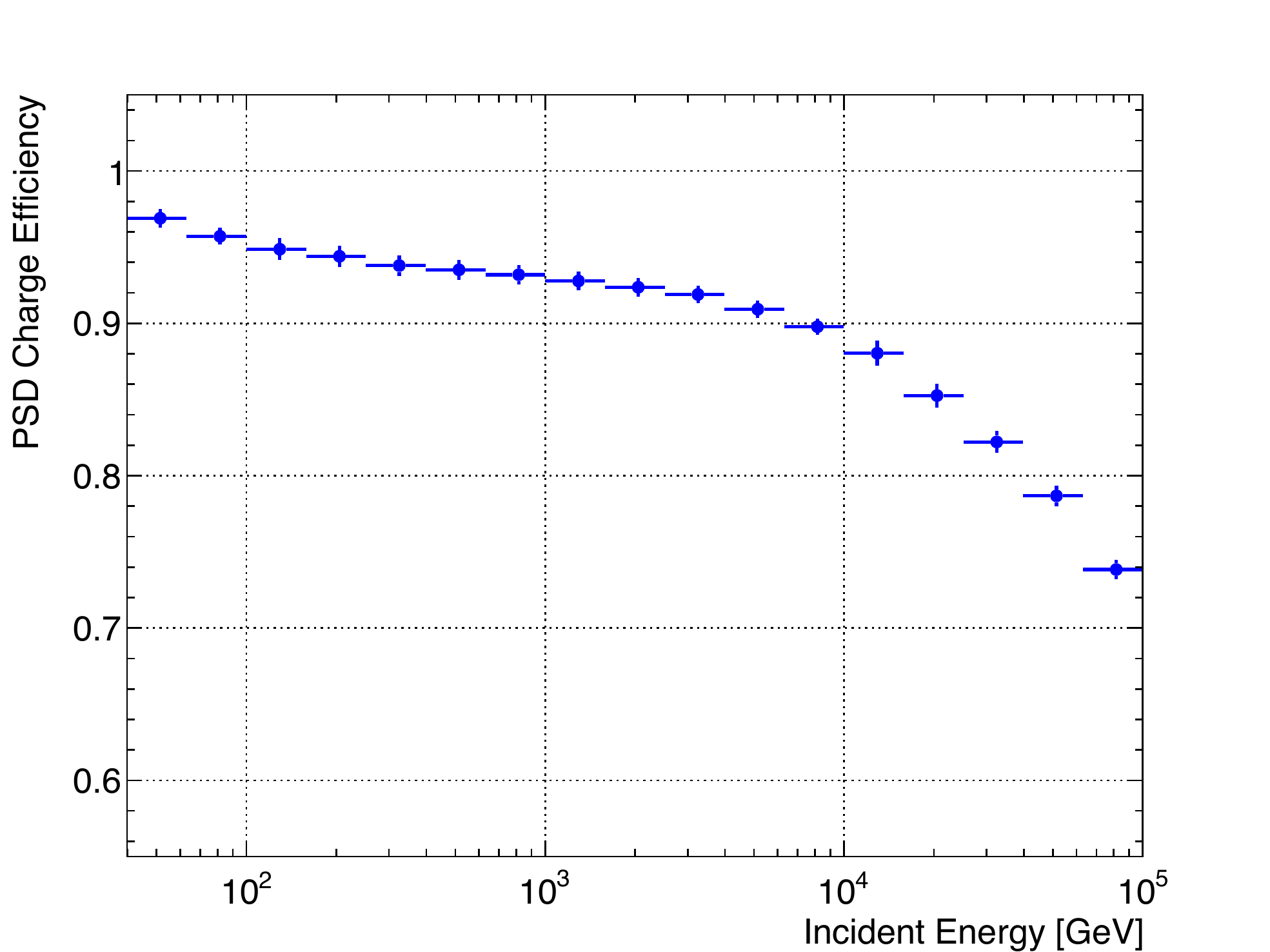}
\put(-40,21){\color{black}{\bf (a)}}
\includegraphics[width=0.5\textwidth]{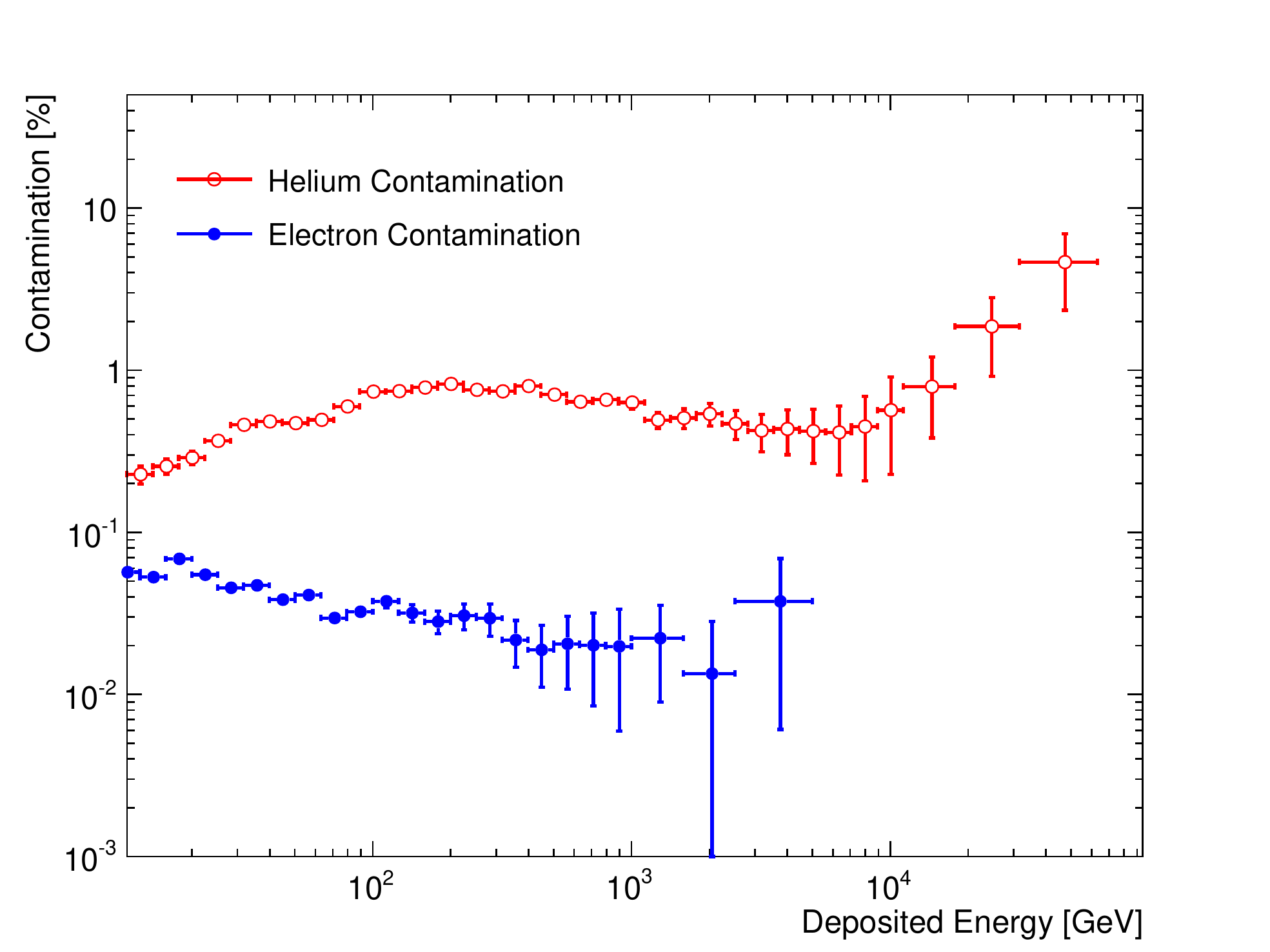}
\put(-40,21){\color{black}{\bf (b)}}
\\
\includegraphics[width=0.5\textwidth]{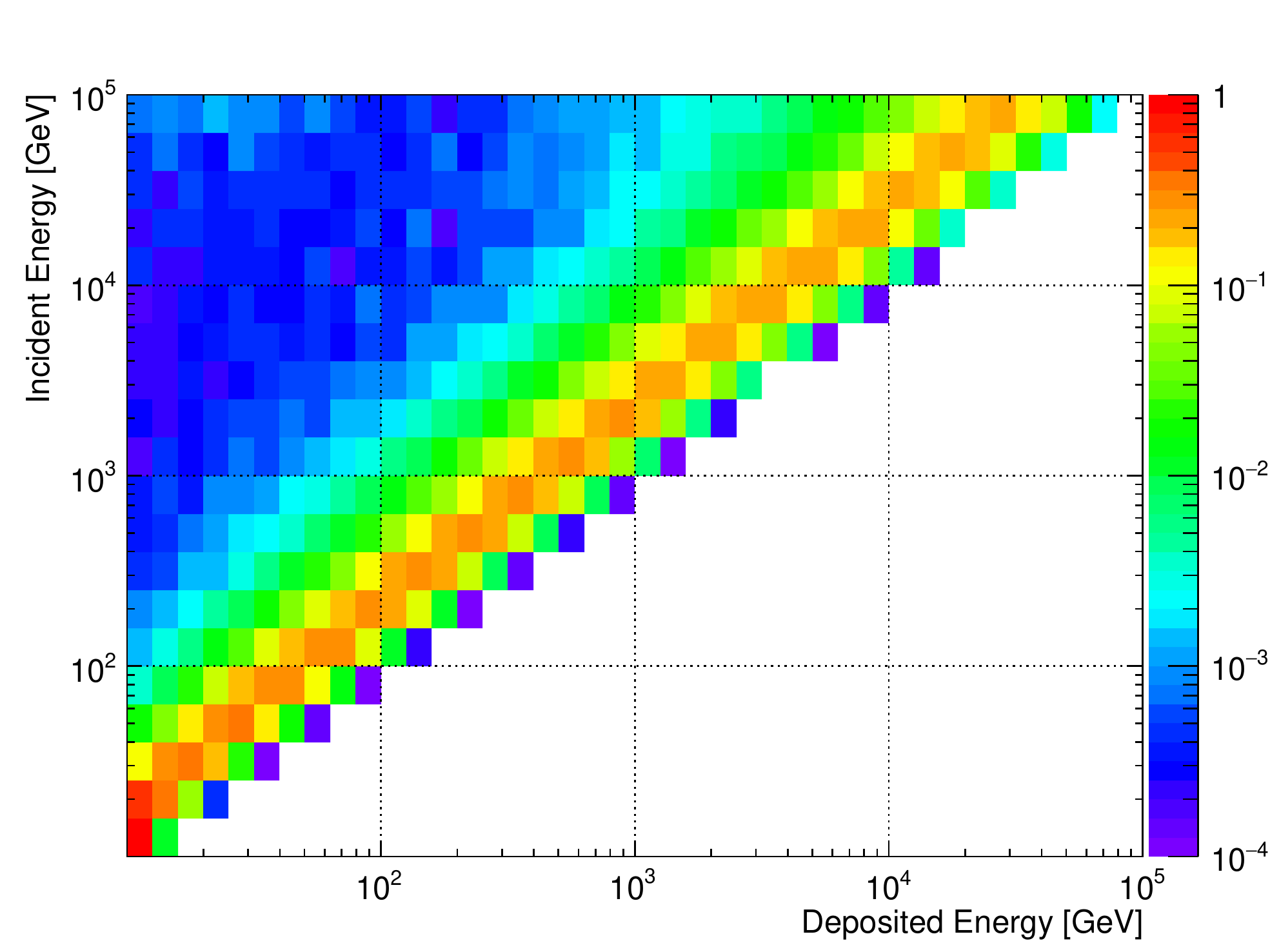}
\put(-40,21){\color{black}{\bf (c)}}
\includegraphics[width=0.5\textwidth]{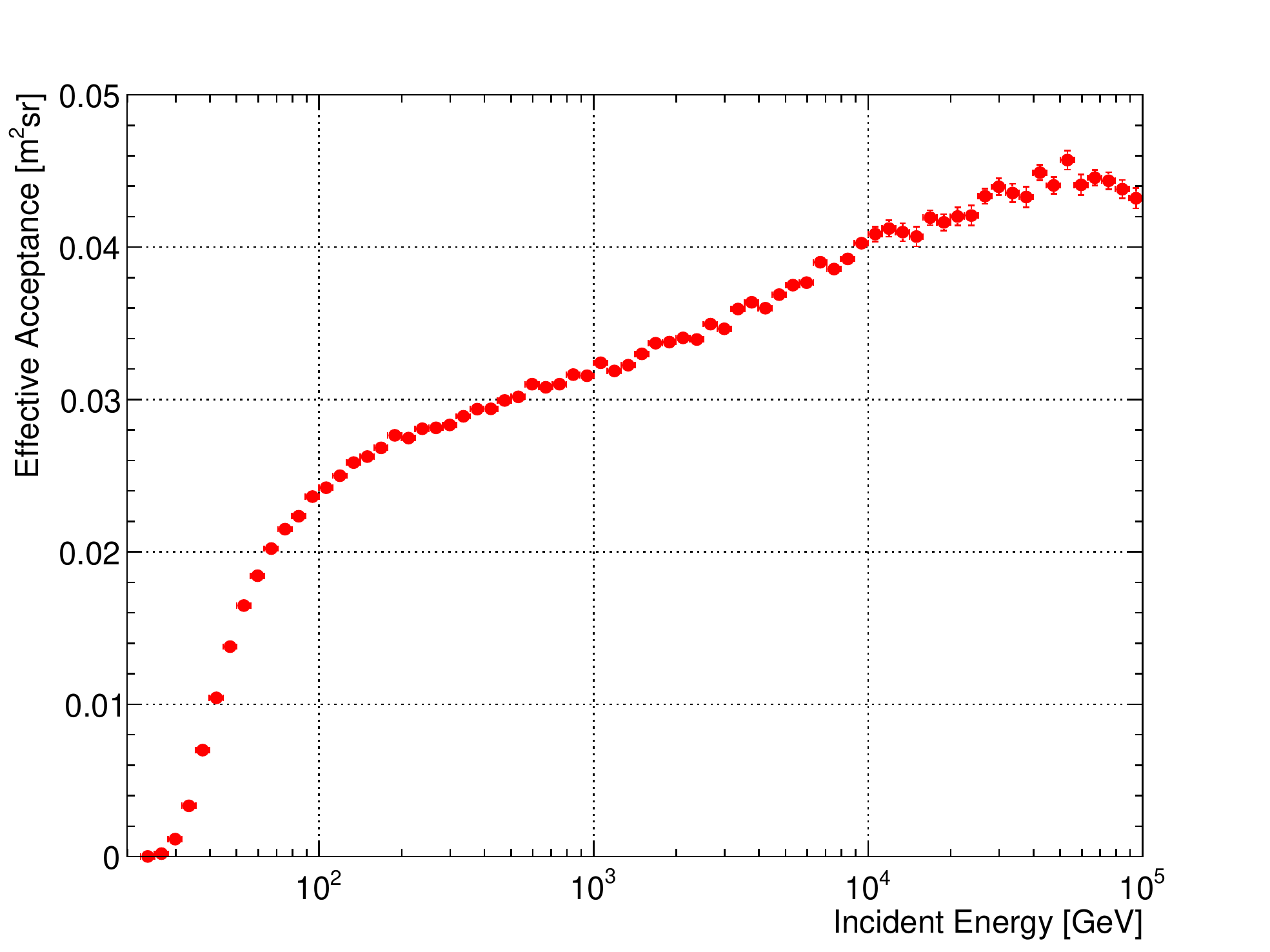}
\put(-40,21){\color{black}{\bf (d)}}
\caption{{\bf Some key information for the proton spectrum measurement.}
{\bf (a)} The charge selection efficiency of protons
versus incident energies for the GEANT FTFP\_BERT model.
{\bf (b)} The fraction of helium (red open circles) and electron
(blue filled dots) backgrounds in the proton candidate events as a
function of deposited energy.
{\bf (c)} Probability distribution of deposited energies in the
BGO calorimeter for different incident energies, for the GEANT FTFP\_BERT
model. The color represents the fraction of events in each energy bin.
{\bf (d)} Effective acceptance of protons versus incident energies
for the GEANT FTFP\_BERT model.
}
\label{fig:Figure3}
\end{figure}

\begin{itemize}

\item Pre-selection\\
DAMPE has four different triggers implemented on orbit: the Unbiased
trigger, the Minimum Ionizing Particle (MIP) trigger, the Low-Energy (LE)
trigger, and the High-Energy (HE) trigger\cite{2017APh....95....6C}.
The events are required to satisfy the HE trigger condition to guarantee
that the shower development starts before or at the top of the calorimeter.
We further require that there are one or more hits in each sub-layer of
the PSD and at least one good track (defined below) in the STK.

To check the MC trigger efficiency with the flight data, the events
with coincidence of signals from the first two BGO layers, tagged as
unbiased triggers, are used. The unbiased trigger events are pre-scaled
by 1/512 at low latitudes ($\le 20^{\circ}$) and 1/2048 at high latitudes.
The HE trigger efficiency is estimated as
$\varepsilon_{\rm trigger} = \frac{N_{\rm HE \mid Unb}}{N_{\rm Unb}}$,
where $N_{\rm Unb}$ is the number of events which pass the unbiased trigger
condition and the proton selection (without the requirement of the HE
trigger), and $N_{\rm HE \mid Unb}$ is the number of events which further
pass the HE trigger.

\item Track selection\\
The track reconstruction algorithm may give more than one track
due to the back-scattering particles. Here we define the ``good'' track
as that the track has at least 4 hits in both the $xz$ and $yz$ layers,
the reduced $\chi^2$ value of the fit is smaller than 25, and the
angular deviation from the BGO shower axis is within $5^{\circ}$.
In the presence of multiple good tracks in the STK, we select the ``best''
one through a combined assessment of the length of the track and the
match between the candidate track and the shower axis in the calorimeter.
Specifically, the selected tracks are required to be the longest
among all the good tracks, among which the one closest to the shower
axis is finally selected.
We then apply the geometry cut on the selected track, which is required
to pass through all the sub-layers of the PSD and the calorimeter from
top to bottom.

To validate the STK track efficiency, we select a proton sample based on
the BGO-reconstructed tracks. The track efficiency is estimated as the
ratio of the number of events passing the above STK track selection
to the total number of events.

\item Charge selection\\
The ionization energy loss in the PSD is employed to measure the charge
of the incoming particle. The charge is measured independently by the
two PSD layers, which are averaged to get the final value.
The measured charge is corrected for the path length of the particle
in the PSD bars using the track information. The PSD and STK alignments
are performed to maximally profit the charge reconstruction capabilities
of the detector\cite{2018arXiv180805720M,2018NIMPA.893...43T}.
The charge measurements of the MC simulations show an energy-dependent
difference from that of the flight data, due primarily to the
back-scattering particles. To reduce the effect on the charge selection
efficiency and the Helium background estimate, we apply energy-dependent
corrections of the charge measurements for the MC simulations. We first
parameterize the charge distributions of protons and helium nuclei in
different deposited energy bins with a Landau-Gaussian convolution
function, and fit the function to the flight data and MC data separately.
Then the MC charges are shifted and shrinked according to the best-fitting
parameters to match with the flight data.
Proton candidates are selected through a cut of the PSD charge.
This cut depends on the BGO deposited energy ($E_{\rm dep}$) as
\begin{equation}
0.6 + 0.05 \cdot \log(E_{\rm dep}/10~{\rm GeV}) \leq Z_{\rm PSD} \leq
1.8 + 0.002 \cdot \log^4(E_{\rm dep}/10~{\rm GeV}).
\end{equation}
Note that this charge selection follows generally the logarithmic
dependence of the ionization energy loss in the PSD with particle energy.
It enables us to have a very small contamination from Helium, i.e.,
less than $1\%$ for deposited energies below 10 TeV and about $5\%$
around 50 TeV. However, this selection results in an energy-dependent
selection efficiency for protons, ranging from $94\%$ for incident energies
of 0.5 TeV to about $75\%$ above 50 TeV. The charge selection efficiency
for protons as a function of incident energy, derived from the GEANT4
FTFP\_BERT MC simulations, is shown in Figure~\ref{fig:Figure3}(a).
This efficiency is used in deriving the effective acceptance shown
in Figure~\ref{fig:Figure3}(d).

The MC charge selection efficiency of each layer of the PSD is also
validated by the flight data. For instance, to estimate the efficiency
of the first PSD layer, we use the second PSD layer and the first STK layer
measurements of the charge to select the sample, and calculate how many of
the events have correct charge measurement in the first PSD layer.
The differences between MC simulations and the flight data are adopted
as systematic uncertainties of the charge selection efficiencies.

\end{itemize}

\subsection*{Background estimate}

The background for protons includes mis-identified helium nuclei and a
tiny fraction of electrons. The electrons are rejected thanks to different
developments between hadronic showers and electromagnetic ones in the
BGO calorimeter. The fraction of residual electrons in the proton sample
is estimated to be about $0.05\%$ for deposited energies larger than
$20$ GeV, using the template fit of the shower morphology parameter
($\zeta$ as defined in Ref.\cite{2017Natur.552...63D}).
Helium nuclei are the main source of background for protons. We employ
the template fit of the PSD charge spectra to estimate the helium
backgrounds. The templates are built based on MC simulations (see
Figure~\ref{fig:Figure1}). The fraction of helium contamination as a
function of deposited energy in the BGO calorimeter is shown in
Figure~\ref{fig:Figure3}(b). It is $\lesssim 1\%$ for deposited energies
below 10 TeV, and increases up to $\sim 5\%$ around $50$ TeV.

\subsection*{Energy measurement and spectral deconvolution}

The energy of an incident particle is measured by the BGO calorimeter.
Due to the limited thickness of the BGO calorimeter ($\sim1.6$
nuclear interaction length) and the missing energy due to muon and
neutrino components in hadronic showers, the energy measurements of CR
nuclei are biased with some uncertainties. Therefore MC simulations
are required to estimate the energy response of the calorimeter.
The energy resolution for protons is found to be about $25\%\sim35\%$
for incident energies from 100 GeV to 10 TeV\cite{2017APh....95....6C}.
The linear region of the energy measurement can extend to incident energies
of $\sim 100$ TeV, thanks to the maximum reachable energy of $4$ TeV for
the dynode-2 readout device of each BGO bar\cite{2017APh....95....6C}.
For very few highest energy events, the saturation of the maximum
energy deposited BGO bar has been corrected based on the simulated
transverse shower profile.
The Engineering Qualification Model of DAMPE was extensively tested using
test beams at the European Organization for Nuclear Research (CERN) in
2014-2015\cite{2016NIMPA.836...98Z,2017NIMPA.856...11Y}. The test beam
momenta are 5, 10, 150, and 400 GeV/c, respectively. The comparison
between the test beam data and MC simulations shows a good agreement
with each other\cite{2017APh....95....6C}. The on-orbit calibration of
the energy measurement is done by means of the MIP signals in each BGO
crystal\cite{DAMPE-Calibration}.

A deconvolution of the measured energy distribution into the incident
energy distribution is applied. The number of events in the $i$-th
deposited energy bin, $N_{{\rm dep},i}$, can be obtained via the sum
of number of events $N_{{\rm inc},j}$ in all the incident energy bins
weighted by the energy response matrix
\begin{equation}
N_{{\rm dep},i}= \sum_{j} M_{ij}N_{{\rm inc},j},
\label{eq-unfold}
\end{equation}
where $M_{ij}$ is the probability that an event in the $j$th incident
energy bin is detected in the $i$-th deposited energy bin. We use MC
simulations to derive the energy response matrix, applying the same
selections as described above. Figure~\ref{fig:Figure3}(c)
shows the energy response matrix for different incident energies, for
the FTFP\_BERT model. The color represents the relative probability
that a proton with $E_{\rm inc}$ deposits $E_{\rm dep}$ energy in the
calorimeter. Eq.~(2) is solved with a Bayesian method to derive the
incident event distribution\cite{1995NIMPA.362..487D}.

\subsection*{Acceptance and absolute flux}

The effective acceptance is defined as the product of the geometric factor
and selection efficiencies (including energy, trigger, track, and charge
selections). The effective acceptance for the $i$-th incident energy bin
is calculated as
\begin{equation}
A_{{\rm eff},i} = A_{\rm gen} \times \frac{N_{{\rm pass},i}}{N_{{\rm gen},i}},
\label{eq-effacc}
\end{equation}
where $A_{\rm gen}$ is the geometrical factor of the MC generation sphere,
$N_{{\rm gen},i}$ and $N_{{\rm pass},i}$ are the numbers of generated events
and those passing the selections.
All efficiencies and the effective acceptance are obtained via the
MC simulations. For the selection efficiencies, we have compared the MC
simulations and the flight data with selected control samples (see the
previous ``Proton event selection" section), and the differences are
adopted as an estimate of the systematic uncertainties of the effective
acceptance. Figure~\ref{fig:Figure3}(d) shows the effective acceptance
as a function of incident energy.

The absolute proton flux $F$ in the incident energy bin
$[E_i,\,E_i+\Delta E_i]$ is then
\begin{equation}
F(E_i,E_i+\Delta E_i) = \frac{N_{{\rm inc},i}}{\Delta E_i~A_{{\rm eff},i}
~T_{\rm exp}},
\label{eq-flux}
\end{equation}
where $\Delta E_i$ is the width of the energy bin and $T_{\rm exp}$ is the
exposure time.

\subsection*{Systematic uncertainties}

Several types of systematic uncertainties are investigated in this analysis,
including the event selection, the background subtraction, the spectral
deconvolution procedure, and the energy response. The systematic
uncertainties related with the event selection are estimated through
comparisons between MC simulations and the flight data. The total uncertainty
of the selection efficiencies is
\begin{equation}
\sigma_{\rm sel} = \sqrt{\sigma_{\rm trigger}^2 + \sigma_{\rm track}^2
+ \sigma_{\rm charge}^2} \approx 4.7\%,
\label{eq-syserr}
\end{equation}
where $\sigma_{\rm trigger} \approx 2.5\%$, $\sigma_{\rm track} \approx
3.5\%$, and $\sigma_{\rm charge} \approx 1.8\%$ are the corresponding
systematic uncertainties of the trigger, track selection, and charge
selection efficiencies.

The uncertainties due to the spectral deconvolution are estimated to
be $\lesssim 1\%$, through re-generation of the response matrix and
varying the spectral index from 2.5 to 3.1 when re-weighting the
simulation data. The systematic uncertainties due to the Helium
background subtraction are estimated through varying the charge
selection condition Eq.~(1) by $\pm10\%$, and repeating the analysis.
The background subtraction gives $\sim0.1\%$ systematic uncertainties
below 40 TeV and increases to $\sim5\%$ at higher energies.
The above systematic uncertainties are added in quadrature to give the
total systematic uncertainties associated with the analysis procedure
($\sigma_{\rm ana}$ as given in Table 1).

The uncertainties of the fluxes due to different hadronic interaction
models are estimated to be about $7\%$ for energies less than 400 GeV
via comparisons of the HE trigger efficiency and the energy
deposit fraction between the 400 GeV test beam data and the
GEANT FTFP\_BERT simulation. Note that the 5 and 10 GeV test beam
data are out of the interested energy range of the current study, and
the 150 GeV test beam is limited by statistics and thus not used.
For higher energies, we use the difference between the GEANT FTFP\_BERT
model and the FLUKA model to estimate such systematic uncertainties,
which vary from $7\%$ to $10\%$. A further check of the DPMJET model 
with the CRMC\cite{crmc-web} 
interface gives negligible difference compared with the FLUKA model.
Finally, the uncertainties associated with the absolute energy scale
are about $2\%$\cite{Zang2017}, which are not corrected in this work.

\begin{figure}[!ht]
\includegraphics[width=1.0\textwidth,angle=0]{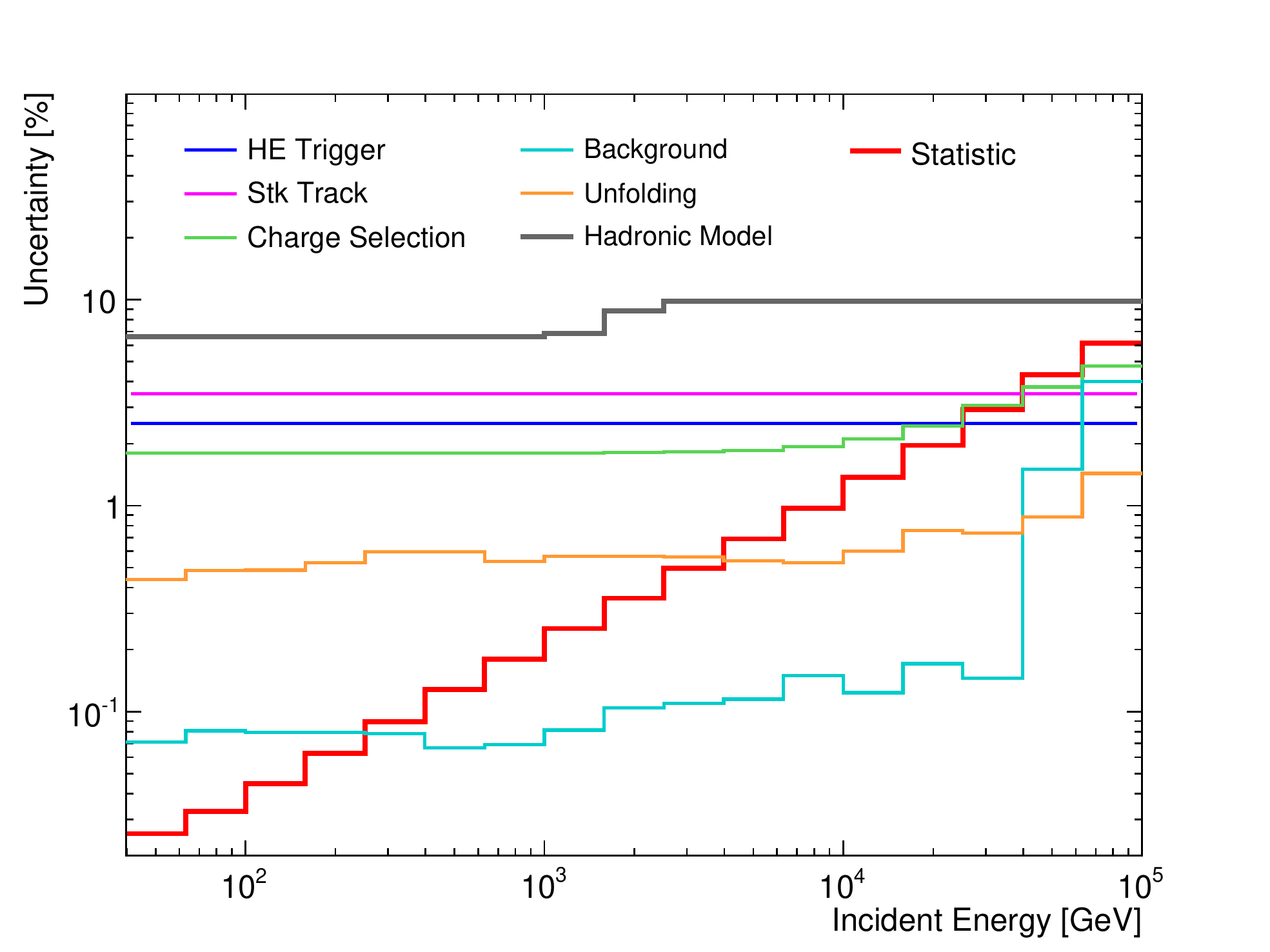}
\caption{{\bf Statistical and systematic uncertainties of the proton
flux measurements.}}
\label{fig:Figure4}
\end{figure}

Figure~\ref{fig:Figure4} summarizes the energy-dependent relative
uncertainties of the proton fluxes.
The statistical uncertainties refer to the Possion fluctuations
of the detected numbers of events in each deposited energy bin. To get
the statistical uncertainties of the unfolded proton fluxes, an error
propagation from the detected events to the unfolded fluxes is necessary
in order to properly take into account the bin-by-bin migration due to
the unfolding procedure\cite{1995NIMPA.362..487D}. To obtain a proper
estimate of the full error propagation, we run toy-MC simulations to
generate fake observations in each deposited energy bin following the
Poisson distribution, and get the proton fluxes through the unfolding
procedure. The root-mean-squares of the resulting proton fluxes are taken
as the $1\sigma$ statistical uncertainties.
We find that the systematic uncertainties due to the
hadronic interaction models dominate in the whole energy range.
The STK track efficiency uncertainties are the sub-dominant ones.
In Table 1 the statistical uncertainties, the systematic one associated
with the analysis procedure (added quadratically), and the systematic
ones due to hadronic models are presented separately.

\subsection*{Comparison of different spectral models}

To quantify the spectral behaviors of the proton spectrum, we use a
power-law (PL) function
\begin{equation}
F(E)=F_0\left(\frac{E}{\rm TeV}\right)^{-\gamma}
\end{equation}
or a smoothly broken power-law (SBPL) one
\begin{equation}
F(E)=F_0\left(\frac{E}{\rm TeV}\right)^{-\gamma}
\left[1+\left(\frac{E}{E_b}\right)^s\right]^{\Delta\gamma/s}
\end{equation}
to fit the data.

\begin{figure}[!ht]
\includegraphics[width=1.0\textwidth,angle=0]{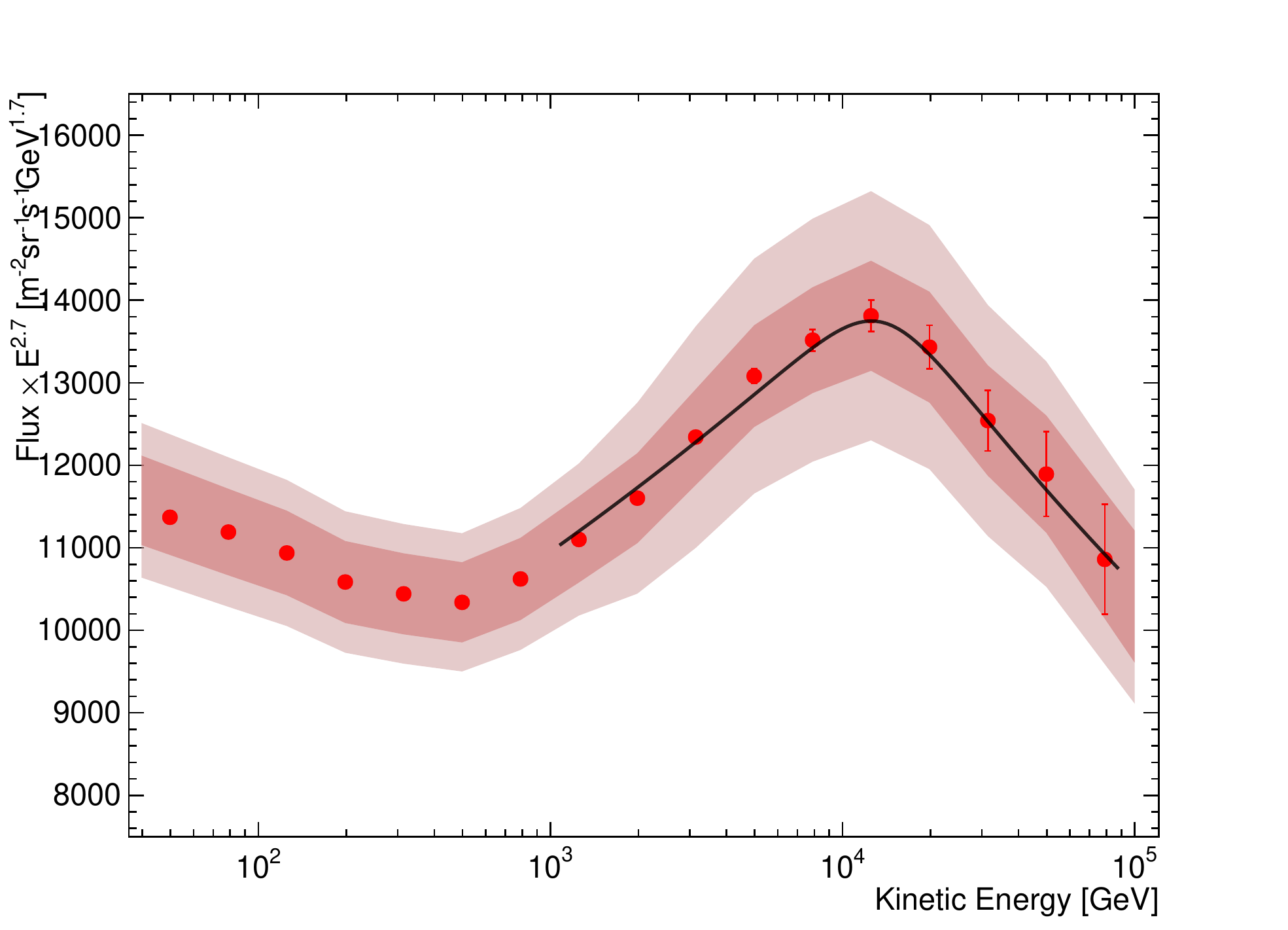}
\caption{{\bf Comparison between the best-fitting of the proton spectrum from
1 TeV to 100 TeV with the smoothly broken power-law function (solid line) and the DAMPE data.}
}
\label{fig:Figure5}
\end{figure}

We first focus on the high-energy part of the spectrum from 1 TeV to
100 TeV. To properly account for the systematic uncertainties, we adopt
a set of independent nuisance parameters $w_j$, which are multiplied
on the input model\cite{2017PhRvD..95h2007A}. The $\chi^2$ function is
defined as
\begin{equation}
\chi^2=\sum_{i=8}^{17}\left[\frac{F(E_{i})S(E_i;\,\boldsymbol{w})-F_i}
{\sigma_{\rm stat,i}}\right]^2 + \sum_{j=1}^{m} \left(\frac{1-w_j}
{\tilde{\sigma}_{\rm sys,j}}\right)^2,
\end{equation}
where $E_i$, $F_i$ and $\sigma_{\rm stat,i}$ are the median energy,
flux and statistical uncertainty of the measurement in the $i$th energy
bin, $F(E_i)$ is the model predicted flux in corresponding energy bin,
$S(E_i;\,\boldsymbol{w})$ is a piecewise function defined by its value $w_j$
in corresponding energy range covered by the $j$th nuisance parameter, and
$\tilde{\sigma}_{\rm sys,j}=\sqrt{\sigma_{\rm ana}^2+\sigma_{\rm had}^2}/F$
is the relative systematic uncertainty of the data in such an energy range.
The last term in the right-hand-side is a Gaussian prior of the nuisance
parameters. The energy range of $[1,100]$ TeV is logarithmically divided
into $m$ pieces. According to the energy-dependence of the systematic
uncertainties in our work (Figure \ref{fig:Figure5}), we adopt $m=4$ in
the energy band of the fit which corresponds to 2 nuisance parameters per
decade of energies.

The fit to the PL model gives $\chi^2/{\rm dof}=28.6/4$, where dof
means the number of degree-of-freedom. For the SBPL model, we fix the
smoothness parameter $s$ to be $5.0$ due to a lack of good constraint
on it, and get $\chi^2/{\rm dof}=2.5/2$. The results differ little for
different values of $s$. The reduction of the $\chi^2$ value is about
26.1 for two more free parameters, suggesting a significance of
$\sim4.7\sigma$ in favor of the SBPL model compared with the PL model.
The fit of the SBPL model gives $F_0=(8.68^{+0.50}_{-0.45})\times 10^{-5}$
GeV$^{-1}$~m$^{-2}$~s$^{-1}$~sr$^{-1}$, $\gamma=2.60\pm0.01$,
$E_b=13.6^{+4.1}_{-4.8}$ TeV, and $\Delta\gamma=-0.25\pm0.07$.
The comparison of the best-fit result (solid line) with the data is
shown in Figure~\ref{fig:Figure5}.
We also test the fitting with $m=3$ (or $5$), and find that the
fitting parameters change very little, while the significance in favor
of the SBPL model becomes $6.9\sigma$ (or $4.2\sigma$).

We also fit the low energy part of the spectrum from 100 GeV to 6.3 TeV
using the SBPL model to address the spectral hardening feature. Again 4
nuisance parameters are assumed, and $s$ is fixed to be $5.0$ which is close
to that obtained by fitting the AMS-02 spectrum\cite{2015PhRvL.114q1103A}.
The fitting model parameters are $F_0=(7.58^{+0.36}_{-0.31})\times 10^{-5}$
GeV$^{-1}$~m$^{-2}$~s$^{-1}$~sr$^{-1}$, $\gamma=2.772\pm0.002$,
$E_b=0.48\pm0.01$ TeV, $\Delta\gamma=0.173\pm0.007$. As comparisons,
the fit to the PAMELA data gives\cite{2011Sci...332...69A}
$E_b=0.232^{+0.035}_{-0.030}$ TeV, $\gamma=2.850\pm0.016$, and
$\Delta\gamma=0.18\pm0.06$, and the fit to the AMS-02 spectrum gives
\cite{2015PhRvL.114q1103A} $E_b=0.34^{+0.09}_{-0.05}$ TeV,
$\gamma=2.849^{+0.006}_{-0.005}$, and $\Delta\gamma=0.133^{+0.056}_{-0.037}$,
respectively. Our low-energy spectrum is slightly harder than that measured
by PAMELA and AMS-02. The break energy inferred from the DAMPE data is
roughly consistent with that of AMS-02, but is slightly higher than that
of PAMELA. The $\Delta\gamma$ values of these three results are consistent
with each other. Nevertheless, we should note that the fitting
functions and energy ranges adopted in Refs.\cite{2011Sci...332...69A,
2015PhRvL.114q1103A} are not exactly the same as those in our work,
and the comparison of the detailed numbers should be done with caution.

{\it Note added:} During the final stage for the publication of this
paper, the CALET collaboration reported new measurements of the proton
spectrum from 50 GeV to 10 TeV and confirmed the spectral hardening feature
found by other measurements\cite{2019PhRvL.122r1102A}. The lack of measurements
above 10 TeV of CALET, however, hampers a crosscheck of our main finding.

\noindent

\newcounter{firstbib}


\section*{Acknowledgements}
The DAMPE mission is funded by the strategic priority science and
technology projects in space science of Chinese Academy of Sciences.

\section*{Funding}
In China the data analysis was supported in part by the National Key
Research and Development Program of China (No. 2016YFA0400200), the
National Natural Science Foundation of China (Nos. 11525313, 11622327,
11722328, U1738205, U1738207, U1738208), the strategic priority science
and technology projects of Chinese Academy of Sciences (No. XDA15051100),
the 100 Talents Program of Chinese Academy of Sciences, and the Young
Elite Scientists Sponsorship Program.
In Europe the activities and the data analysis are supported by the
Swiss National Science Foundation (SNSF), Switzerland; the National
Institute for Nuclear Physics (INFN), Italy.

\section*{Author contributions}
This work is the result of the contributions and efforts of all the
participating institutes. All authors have reviewed, discussed, and
commented on the results and on the manuscript. In line with the
collaboration policy, the authors are listed here alphabetically.

\section*{Competing interests}
The authors declare no competing interests.

\section*{Data and materials availability}
The cosmic ray proton fluxes along with statistical and systematics
uncertainties are available in Table 1. Additional requests can be
addressed to the DAMPE Collaboration (dampe@pmo.ac.cn).

\newpage

\vfill

\vfill

\begin{table}
\begin{center}
\title{}Table 1. {\bf Fluxes of CR protons measured with DAMPE,
together with 1$\sigma$ statistical and systematic uncertainties.}
The systematic uncertainties include those associated with the analysis
procedure $\sigma_{\rm ana}$ (e.g., the event selection, the background
subtraction, and the spectral deconvolution), and the energy responses
due to different hadronic models $\sigma_{\rm had}$.\\
\begin{tabular}{ccccc}
\hline
$\langle E \rangle$ & $E_{\rm min}$ & $E_{\rm max}$ & $F \pm \sigma_{\rm stat} \pm \sigma_{\rm ana} \pm \sigma_{\rm had}$\\
(GeV) & (GeV) & (GeV) & (GeV$^{-1}$m$^{-2}$s$^{-1}$sr$^{-1}$) \\
\hline
49.8 & 39.8 & 63.1 & $(2.97 \pm 0.00 \pm 0.14 \pm 0.20)\times 10^{-1}$ \\
78.9 & 63.1 & 100.0 & $(8.43 \pm 0.00 \pm 0.40 \pm 0.56)\times 10^{-2}$ \\
125.1 & 100.0 & 158.5 & $(2.38 \pm 0.00 \pm 0.11 \pm 0.16)\times 10^{-2}$ \\
198.3 & 158.5 & 251.2 & $(6.64 \pm 0.00 \pm 0.31 \pm 0.44)\times 10^{-3}$ \\
314.3 & 251.2 & 398.1 & $(1.89 \pm 0.00 \pm 0.09 \pm 0.12)\times 10^{-3}$ \\
498.1 & 398.1 & 631.0 & $(5.39 \pm 0.01 \pm 0.25 \pm 0.36)\times 10^{-4}$ \\
789.5 & 631.0 & 1000 & $(1.60 \pm 0.00 \pm 0.07 \pm 0.11)\times 10^{-4}$ \\
1251 & 1000 & 1585 & $(4.81 \pm 0.01 \pm 0.23 \pm 0.33)\times 10^{-5}$ \\
1983 & 1585 & 2512 & $(1.45 \pm 0.01 \pm 0.07 \pm 0.13)\times 10^{-5}$ \\
3143 & 2512 & 3981 & $(4.45 \pm 0.02 \pm 0.21 \pm 0.44)\times 10^{-6}$ \\
4981 & 3981 & 6310 & $(1.36 \pm 0.01 \pm 0.06 \pm 0.13)\times 10^{-6}$ \\
7895 & 6310 & 10000 & $(4.06 \pm 0.04 \pm 0.19 \pm 0.40)\times 10^{-7}$ \\
12512 & 10000 & 15849 & $(1.20 \pm 0.02 \pm 0.06 \pm 0.12)\times 10^{-7}$ \\
19830 & 15849 & 25119 & $(3.35 \pm 0.07 \pm 0.17 \pm 0.33)\times 10^{-8}$ \\
31429 & 25119 & 39811 & $(9.03 \pm 0.26 \pm 0.48 \pm 0.89)\times 10^{-9}$ \\
49812 & 39811 & 63096 & $(2.47 \pm 0.11 \pm 0.15 \pm 0.24)\times 10^{-9}$ \\
78946 & 63096 & 100000 & $(6.50 \pm 0.40 \pm 0.50 \pm 0.64)\times 10^{-10}$ \\
\hline
\end{tabular}
\label{tab:Table1}
\end{center}
\end{table}

\vfill

\vfill

\vfill

\clearpage

\end{document}